\documentclass[aps,prl,a4paper,twocolumn,amsmath,amssymb,amsfonts,nofootinbib,floatfix,superscriptaddress]{revtex4-1}
\pdfoutput=1
\usepackage{graphicx} 
\usepackage[usenames, dvipsnames]{color}
\usepackage{dcolumn} 
\usepackage{bm}
\usepackage{undertilde}
\usepackage{soul}
\usepackage{accents}
\usepackage{enumitem}
\usepackage{hyperref}
\usepackage{lipsum}
\usepackage[utf8]{inputenc}
\usepackage{natbib}
\usepackage{bookmark}

\hypersetup{
    citecolor=OliveGreen,
    colorlinks=true,
    urlcolor=blue
}

\renewcommand{\SS}{S} 

\newcommand{\z}{\mathbf{z}}

\newcommand{\CB}{\mathbb{C}}
\newcommand{\FC}{\mathcal{F}}
\newcommand{\GC}{\mathcal{G}}
\newcommand{\HB}{\mathbb{H}}
\newcommand{\HC}{\mathcal{H}}
\newcommand{\QB}{\mathbb{Q}}
\newcommand{\QC}{\mathcal{Q}}
\newcommand{\BG}{\text{BG}}
\newcommand{\Hopt}{H_o}

\let\OldSection=\section
\renewcommand{\section}[1]{\emph{#1}.~}

\definecolor{red}{rgb}{0.7,0.1,0.1}
\definecolor{Red}{rgb}{1,0,0}
\definecolor{green}{rgb}{0,0.8,0}
\definecolor{blue}{rgb}{0,0,0.8}
\definecolor{cautionred}{rgb}{1.0,0,0}
\definecolor{maroon}{rgb}{0.7,0,0}
\definecolor{ngreen}{rgb}{0.3,0.7,0.3}
\definecolor{golden}{rgb}{0.8,0.6,0.1}

\newcommand{\blk}{\color{black}}

\newcommand{\fref}[1]{Fig.~(\ref{#1})}
\newcommand{\eref}[1]{Eq.~(\ref{#1})}
\newcommand{\tref}[1]{Table~\ref{#1}}
\newcommand{\beq}{\begin{align}}
\newcommand{\eeq}{\end{align}}
\newcommand{\beqa}{\begin{align}}
\newcommand{\eeqa}{\end{align}}
\newcommand{\beqan}{\begin{align*}}
\newcommand{\eeqan}{\end{align*}}

\newcommand{\eq}[1]{Eq.~(\ref{#1})}
\newcommand{\eqs}[1]{Eqs.~(\ref{#1})}
\newcommand{\eqr}[1]{(\ref{#1})}

\newcommand{\etal}{ \emph{et al.}}

\newcommand{\specialcell}[2][c]{%
  \begin{tabular}[#1]{@{}c@{}}#2\end{tabular}}

\begin{document}

\title{
Thermodynamics from first principles: correlations and nonextensivity
} 

\author{S.~N.~Saadatmand} \email{n.saadatmand@griffith.edu.au}
\affiliation{Centre for Quantum Dynamics, Griffith University, Nathan, QLD 4111, Australia.
}%
\author{Tim Gould}
\affiliation{Qld Micro- and Nanotechnology Centre, Griffith University, Nathan, QLD 4111, Australia.
}%
\author{E.~G.~Cavalcanti\blk}
\affiliation{Centre for Quantum Dynamics, Griffith University, Gold Coast, QLD 4222, Australia.
}%
\author{J.~A.~Vaccaro}
\affiliation{Centre for Quantum Dynamics, Griffith University, Nathan, QLD 4111, Australia.
}%
\date{\today}

\begin{abstract}
The standard formulation of thermostatistics, being based on the Boltzmann-Gibbs distribution and logarithmic Shannon entropy, describes idealized uncorrelated systems with extensive energies and short-range interactions. In this letter, we use the fundamental principles of ergodicity (via Liouville's theorem), the self-similarity of correlations, and the existence of the thermodynamic limit to derive generalized forms of the equilibrium
distribution for long-range-interacting systems. Significantly, our formalism provides a
justification for the well-studied nonextensive thermostatistics characterized by
the Tsallis distribution, which it includes as a special case.
We also give the complementary maximum entropy derivation of the same distributions by constrained maximization of the Boltzmann-Gibbs-Shannon entropy.
The consistency between the ergodic
and maximum entropy approaches clarifies the use of the latter in the study of correlations and nonextensive
thermodynamics.
\end{abstract}

\maketitle

\section{Introduction}
The ability to describe the statistical state of a macroscopic system is central to many areas of physics~\protect\cite{Reif1965,Abe2001c,Naudts2011,Tsallis2009}.
In thermostatistics, the statistical state of a system of $N$ particles in equilibrium is described by the distribution function $w_{\z}$ over $\z$ where   $\z=(\{q_1,\cdots,q_N\},\{p_1,\cdots,p_N\})$ \blk defines a
point (the microstate) in the concomitant $6N$-dimensional phase space.
The central question addressed in this letter
is, \emph{what is the generalized form of $w_{\z}$
for a composite
system at thermodynamic equilibrium that features correlated subsystems?}

This question has been the subject of intense research for more than a
century~\protect\cite{Gibbs1902,Jaynes1957,Jaynes1957a,Tsallis1988,Plastino1994,Tsallis1998,Martinez2001,Abe2001,Dauxois2002,Plastino2004,Abe2006,Suyari2006,Naudts2011,Caruso2008,Tsallis2009,Tsallis2009a,Tsallis2011,Akkermans2012,Deppman2016,Oikonomou2017,Jizba2019}.
Correlations and nonextensive energies are associated with long-range-interacting systems, which are at the focus of much of the effort (in particular see~\protect\cite{Plastino2004,Akkermans2012,Deppman2016,Tsallis2011,Jizba2019}).
The most widespread approach to finding $w_{\z}$ is
the   maximum entropy \blk (MaxEnt) principle
introduced by Jaynes~\protect\cite{Jaynes1957,Jaynes1957a} on the basis of information theory.  The principle entails making the \emph{least-biased} statistical inferences about a physical system consistent with prior 
expected values of a set of its quantities $\{\bar{f}^{(1)},\bar{f}^{(2)},\ldots,\bar{f}^{(l)}\}$.
It requires the distribution $w_{\z}$ to maximize the Gibbs-Shannon (GS) logarithmic entropy functional  $\SS^{\text{GS}}(\{w_{\z}\})$ subject to constraints
$\sum_{\z} f_{\z}^{(i)} w_{\z}=\bar{f}^{(i)}$. Here,
$\SS^{\text{GS}}(\{w_{\z}\}) = -k \sum_{\z} w_{\z} \ln (w_{\z})$
for constant $k>0$.
Despite outstanding success~\protect\cite{Jaynes1983,Presse2013a} in capturing
thermodynamics of weakly-interacting gases,
the principle -- in its original form -- does not describe correlated systems.
Attempts have been made to generalize the principle, however, as there is no accepted method for doing so, controversy has ensued~\protect\cite{Presse2013,Presse2015,Tsallis2015}.

One approach is based on the extension of the MaxEnt principle by Shore and Johnson~\protect\cite{Shore1980}, and entails generalizing the way knowledge of the system is represented by constraints~\protect\cite{Shore1980,Presse2013a}.  Information about correlations are incorporated, e.g.~by modifying the partition function~\protect\cite{Presse2011a} or the structure of the microstates~\protect\cite{Presse2015}.
Another widely-used approach is to generalize the MaxEnt principle   to apply to \blk
a different entropy functional in place of $\SS^{\text{BGS}}(\{w_{\z}\})$.
At the forefront of this effort is the so-called $q$-thermostatistics
based on Tsallis'
entropy $\SS^{\rm Ts}_q(\{w_{\z}\}) \propto (1-\sum_{\z}w_{\z}^q)/(q-1)$ and expressing the constraints as averages with respect to escort probabilities $\{w_{\z}^q\}$~\protect\cite{Tsallis1988,Tsallis2009}.
$q$-thermostatistics is
known to describe a wide range of physical
scenarios~\cite{Uys2001,Beck2000,Beck2001,Boghosian1996,Anteneodo1997,Lima2000,Dauxois2002,Naudts2011,Tirnakli2000,Tirnakli2002,Lavagno1998,Plastino2004,Latora2002,Deppman2016,Zborovsky2018,Plastino1995,Frank1999,Salazar1999,Salazar2000,Portesi1995},
including high-$T_c$ superconductivity, long-range-interacting Ising magnets,
turbulent pure-electron plasmas, $N$-body self-gravitating stellar systems, high-energy hadronic collisions,
and low-dimensional chaotic maps.
The approach has also been 
refined and extended~\protect\cite{Abe1997,Borges1998,Landsberg1998,Naudts2006,Naudts2011}.

A contentious issue, however, is that the Tsallis entropy
does not satisfy Shore and Johnson's system-independence axiom \cite{Shore1980,Presse2013,Presse2015,Tsallis2015}.
Although Jizba\etal~\protect\cite{Jizba2019} recently made some headway towards a resolution, objections remain~\protect\cite{Oikonomou2019}, and the generalization of the MaxEnt principle continues to be controversial.
This brings into focus the need for an independent approach to our central question.

We propose an answer by introducing a general formalism based on
ergodicity~\protect\cite{Reif1965}
for deriving equilibrium distributions, including ones for correlated systems.
Previously this derivation was thought impossible as correlations have been linked with nonergodicity (see e.g.~\protect\cite{Tsallis2009}, p.~68
and p.~320). However, we circumvent these difficulties by showing how the self-similarity of correlations can be invoked to derive the distribution $w_{\z}$ under well-defined criteria.
Then we show how to employ the MaxEnt principle consistently with correlations encoded as a self-similarity constraint function.
After comparing our results with previous works, we present a numerical example for completeness and end with a conclusion.

\section{Key ideas}
Our approach rests on two key ideas.

{(i) Liouville's theorem   for equilibrium systems\blk.} Consider
a generic, classical, dynamical system described by Hamiltonian $H$ and phase-space distribution $w(\z;t)$. Being a Hamiltonian system guarantees the incompressibility of phase-space flows, which is represented via Liouville's equation~\protect\cite{Gibbs1902,Reif1965} by $w$ being a constant of motion along a trajectory,  i.e.
\begin{equation}
\label{eq:HamLR}
  \frac{\partial w}{\partial t} + \dot{\z} \cdot \nabla w
  = \frac{\partial w}{\partial t} + \{w,H\}
  = \frac{d w}{d t} = 0,
\end{equation}
where $\{\;,\;\}$ denotes the Poisson bracket. Imposing the equilibrium condition $\partial w/\partial t=0$ implies
\begin{align}
\label{eq:equilibrium}
        \partial w/\partial t = -\{w,H\}=0,
\end{align}
i.e.~the existence of a steady-state, $w(\z;t)=w_{\z}\,\forall\,t$. Any Hamiltonian \emph{ergodic} system  at equilibrium will obey this condition.
A possible solution of \eq{eq:equilibrium} is given by $w_{\z}=\frak{a}f(\frak{b}H_{\z} + \frak{c})$, where $f(\cdot)$ is any differentiable function, for macrostate-defining and normalisation constants $\frak{a}$, $\frak{b}$ and $\frak{c}$ (see e.g.~\protect\cite{SM,Reif1965}). 
We only consider solutions of this form, which is equivalent to invoking the fundamental postulate of equal a priori probabilities for accessible microstates~\cite{Reif1965}.
For brevity, we shall write the solution as 
\begin{align}
    w_\z=\GC_X(H_\z)  \label{eq:w=w(H)}
\end{align} 
and leave the dependence on the parameters $\frak{a}$, $\frak{b}$ and $\frak{c}$ as being implicit in the label $X$.

{(ii)   Deriving equilibrium distributions.}
Consider the equilibrium distributions $w^{A}$, $w^B$ and Hamiltonians $H^A$, $H^B$ of two isolated, conservative, short-range-interacting systems labelled $A$ and $B$ where
\begin{align}    \label{eq:H^AB sum}
    H^{AB}_{\z_{AB}}&=H^A_{\z_{A}}+H^B_{\z_{B}} ,\\
     w^{AB}_{\z_{AB}} &= w^{A}_{\z_{A}}w^{B}_{\z_{B}}    \label{eq:w^AB product}
\end{align}
are the total Hamiltonian and joint distributions, respectively, for the isolated, composite system $AB$ at equilibrium, and  $\z_{AB}\equiv(\z_{A},\z_{B})$.
From \eqr{eq:w=w(H)}, each distribution is a function of its respective Hamiltonian.
Taken together, \eqs{eq:w=w(H)}-\eqr{eq:w^AB product} imply the general solution $w^X_{\z_X}$  is the Boltzmann-Gibbs (BG) distribution $\GC_X(H^X_{\z_X})=\frak{a} e^{\frak{b} H_{\z_X}}$ for macrostate-dependent constants $\frak{a}$, $\frak{b}$ and $X=A, B~\text{and}~AB$.

This well-known result can easily be generalised.  For example, replacing \eq{eq:w^AB product} with
\begin{align}    \label{eq:w^AB q product}
     w^{AB}_{\z_{AB}}&=w^{A}_{\z_{A}}\otimes_q w^B_{\z_{B}},
\end{align}
where $\otimes_q$ is the $q$-product~\cite{Suyari2006}, correspondingly implies that the general solution is given by the Tsallis distribution $\GC^X(H_{\z_X})=\frak{a} e_q^{\frak{b} H_{\z_X}}$ where $e_q^x$ is the $q$-exponential of $x$
provided due care is taken in respect of applying the $q$-algebra~\protect{\cite{Suyari2006,Nelson2008}} and normalisation~\protect{\cite{SM}}. Note that each $w_{\z_X}^X$ is the equilibrium distribution for system $X$ in isolation, and \eq{eq:w^AB q product} represents a correlated state of $A$ and $B$,
where $w_{\z_A}^A$ and $w_{\z_B}^B$ are \emph{not} the marginals of $w_{\z_{AB}}^{AB}$ for $q\ne 1$. 
As this result has previously been regarded~\cite{Tsallis2009} as incompatible with \eq{eq:equilibrium}, it shows that Liouville's theorem has an underappreciated application for describing highly correlated systems.

\section{Finding a generalized distribution}
With these ideas in mind, we derive our main results for a composite, 
self-similar, classical Hamiltonian system in thermodynamic equilibrium.
For brevity, we explicitly treat a composite system $AB$ composed of two subsystems $A$ and $B$, although our results are easily extendable to compositions involving
an arbitrary number of macroscopic subsystems.
Let the tuples $(w^{AB}_{\z_{AB}},H^{AB}_{\z_{AB}})$,
$(w^{A}_{\z_{A}},H^{A}_{\z_{A}})$,
$(w^{B}_{\z_{B}},H^{B}_{\z_{B}})$
denote the composite and isolated equilibrium distributions and
Hamiltonians of the composite $AB$, and separate
$A$, $B$ subsystems, respectively;
$w^X_{\z_X}$ is the equilibrium probability that system $X$ is in phase space point $\z_X$. The following criteria encapsulate properties of the system required for subsequent work. They immediately lead to two key Theorems, which generalize thermostatistics.

\emph{Criterion I -- Thermodynamic limit}:
Consider a sequence of  systems $A_1, A_2, \cdots$ for which the solution \eq{eq:w=w(H)} for the $n$th term is given by $w^{A_n}_{\z_{A_n}}=\GC^{(n)}_{A_n}(H^{A_n}_{\z_{A_n}})$.
A sequence that increases in size is said to have a \emph{thermodynamic limit} if $\GC^{(n)}_{A_n}$ attains a limiting parametrized form as $A_n$ becomes macroscopic, i.e. if $\GC^{(n)}_{A_n}\to \GC_{A}$ as $n\to\infty$. 
The distribution $w^{A}_{\z_{A}}=\GC_{A}(H^{A}_{\z_{A}})$, where the dependence on system, macrostate, and normalisation constants is implicit in the label $A$ on $\GC_A$,
is said to represent the thermostatistical properties of the
physical material comprising $A$ in the thermodynamic limit.

Examples of limiting forms include the BG distribution
$\GC(H_{\z}) = \frak{a}e^{-\frak{b} H_{\z}}$
and the Tsallis distribution
$\GC(H_{\z})= \frak{a}e_q^{-\frak{b} H_{\z}}$
for macrostate-dependent parameter $\frak{b}$ and
normalization constant $\frak{a}$. 

\emph{Criterion II -- Compositional self-similarity}:
We define a  system as having compositional self-similarity
if there exist mapping functions $\CB$ and $\HB$ such that the composite equilibrium distribution and energy of macroscopic $AB$ are related to the isolated equilibrium distribution and energy of macroscopic $A$ and $B$ by the following relations
\begin{subequations}
\label{eq:two-variables}
\begin{align}
\label{eq:two-variable2}
  H^{AB}_{\z_{AB}} &= \HB(H^A_{\z_{A}}, H^B_{\z_{B}})~\\
\label{eq:two-variable1}
  w^{AB}_{\z_{AB}} &= \CB(w^{A}_{\z_{A}},w^{B}_{\z_{B}}),
    ~ 0 \leq \CB \leq 1~, 
\end{align}
\end{subequations}
for all $\z_{AB}$,  where $\HB$ embodies the nature of the interactions, and
$\CB$ embodies the nature of the correlations.
For example, short-range interactions are well approximated by $H^{AB}_{\z_{AB}}=H^A_{\z_{A}}+ H^B_{\z_{B}}$ and $w^{AB}_{\z_{AB}} = w^{A}_{\z_{A}} w^{B}_{\z_{B}}$, whereas  the Tsallis distribution in \eq{eq:w^AB q product} has been applied to a wide range of physical situations \cite{Uys2001,Beck2000,Beck2001,Boghosian1996,Anteneodo1997,Lima2000,Dauxois2002,Naudts2011,Tirnakli2000,Tirnakli2002,Lavagno1998,Plastino2004,Latora2002,Deppman2016,Zborovsky2018,Plastino1995,Frank1999,Salazar1999,Salazar2000,Portesi1995} exhibiting strong correlations and long-range interactions.  Other relations hold in general, as shown in \tref{table:main}  \protect\cite{SM}.
For brevity we will henceforth
use ``self-similar" to refer to compositional self-similarity.

\emph{Theorem I}: For systems satisfying compositional self-similarity in the thermodynamic limit, the equilibrium distribution is given by $w^X_{\z_X}=\GC_X(H^X_{\z_X})$ where the function $\GC_X$  satisfies
\begin{align}
\label{eq:T2}
\CB(\GC_{A}(H^{A}_{\z_{A}}),\GC_{B}(H^{B}_{\z_{B}})) =
      \GC_{AB}(\HB(H^A_{\z_A},H^B_{\z_B})).
\end{align}
\emph{Proof}: This follows directly from Criteria I and II~$\Box$.

Hence, finding a $\GC$ that satisfies \eq{eq:T2} allows one to calculate the equilibrium distribution in \eq{eq:w=w(H)}.
See Supplementary Material~\cite{SM} for a simple example.
In general, finding $\GC$ is difficult, however, the next theorem supplies a solution for an important class of situations.

\newcommand\T{\rule{0pt}{2.1ex}}
\begin{table*}[htb]
  \caption{A summary of appropriate choices for $\{\FC,\HC\}$
  to reproduce well-established classes of thermostatistics, which
  allowed us to also indicate their potential limitations.
  We have included the conventional partition-funtion-type normalization constants in
  some cases for completeness. Note, however, that such constants can be re-expressed as $a$ and $b$ or $\beta$
  and $\Hopt$--- i.e.~$Z_{\text{BG}}=e^{-\beta \Hopt+k^{-1}S^{\rm GS}}$ and $Z_q=e_q^{-\beta_q \Hopt}$
  (where $\beta_q = \beta[1+(1-q)\beta \Hopt]^{-1}$).
  \label{table:main}}
  \begin{center}
    \begin{tabular}{ |c|c|c|c|c|c|c|c| }
        \hline \hline
        \footnotesize \specialcell[c]{Type of\\thermostatistics} \normalsize &
        \footnotesize Correlations\normalsize \T & \footnotesize
        Hamiltonian \normalsize & \footnotesize \specialcell[c]{
        $\FC(w)$} \normalsize & \footnotesize $\HC(H)$ \normalsize 
        &  \footnotesize Distribution \normalsize & 
        \footnotesize \specialcell[c]{Fails to\\describe:} \normalsize \\ \hline
        \footnotesize this work \normalsize  & \footnotesize \specialcell[c]{$\CB(w_1,w_2)$\\(self-similar)} \normalsize &
        \footnotesize \specialcell[c]{$\HB(H_1,H_2)$\\(arbitrary)} \normalsize & - & - & \footnotesize
        \eref{eq:dist} \normalsize & 
        \footnotesize \specialcell[c]{systems failing\\Criteria I and II} \normalsize
        \\ \hline
        \footnotesize \specialcell[c]{conventional\\thermostatistics~\citep{Gibbs1902,Jaynes1957,Jaynes1957a} }
        \normalsize &
        \footnotesize \specialcell[c]{$w_1 w_2$\\(independent)} \normalsize &
        \footnotesize \specialcell[c]{$H_1 + H_2$\\(noninteracting)} \normalsize&
        $\ln(w)$ & \footnotesize $H$ \normalsize 
        & \footnotesize \specialcell[c]{$\frac{1}{Z_{\text{\BG}}}e^{-\beta H}$
        \\(exponential class)} \normalsize &
        \footnotesize
        \specialcell[c]{correlations,\\nonadditive
        \\Hamiltonians} \normalsize \\ \hline
        \footnotesize \specialcell[c]{Tsallis' ($q$-)\\thermostatistics~\citep{Tsallis1988,Tsallis2009}} \normalsize &
        \footnotesize \specialcell[c]{$w_1 \otimes_q w_2$\\(correlated)} \normalsize &
        \footnotesize \specialcell[c]{$H_1 + H_2$\\(noninteracting)} \normalsize&
        $\ln_q(w)$ & \footnotesize $H$ \normalsize 
        &
        \footnotesize \specialcell[c]{$\frac{1}{Z_q}e_q^{-\beta_qH}$
        \\($q$-deformed class)} \normalsize &
        \footnotesize
        \specialcell[c]{nonadditive
        \\Hamiltonians} \normalsize \\ \hline
        \footnotesize \specialcell[c]{an exactly-solvable\\example exhibiting\\both
        correlations and\\nonextensive energies} \normalsize &
        \footnotesize \specialcell[c]{$w_1 \otimes_q w_2$\\(correlated)} \normalsize &
        \footnotesize \specialcell[c]{$H_1 \oplus_p H_2$\\(interacting)} \normalsize &
        $\ln_q(w)$ & \footnotesize $\ln(e_p^H)$ \normalsize & 
        \footnotesize \specialcell[c]{$\exp_q\big(-\beta\ln(e_p^H)$\\$+\Hopt\big)$} \normalsize
        & - \\ \hline
        \hline
    \end{tabular}
  \end{center}
\end{table*}

\emph{Theorem II}:
Given single-variable invertible maps $\FC$ and $\HC$
satisfying the following functional equations
\begin{subequations}
\label{eq:maps}
\begin{align}
    \FC_{AB}(\CB(w^{A}, w^{B})) &= \FC_{A}( w^{A}) + \FC_{B}( w^{B}) 
            \label{eq:maps a}\\
    \HC_{AB}(\HB(H^A, H^B)) &= \HC_{A}(H^A) + \HC_{B}(H^B) 
         \label{eq:maps b}
\end{align}
\end{subequations}
then there exists a family of equilibrium distributions given by
\begin{align}
   w^X_\z \equiv \GC_{X}(H^X_\z) =
     \FC_X^{-1}(a^X\HC(H^X_\z)+b^X)~~\forall\z
\label{eq:dist}
\end{align}
where $a^X$ and $b^X$ are constants obeying the system composition rules
\begin{align}
   \label{eq:a b composition rules}
    a^{AB} =  a^A = a^B,\qquad  b^{AB} = b^A +b^B. 
\end{align} 
Note that $a^X$ and $b^X$ are generalisations of a common inverse-temperature-like quantity $\beta=a^X$ and an extensive average-energy-like quantity $\Hopt^X=-b^X/\beta$ in the more familiar form of \eq{eq:dist}, $w^X=\FC^{-1}(\beta(\HC(H^X)-\Hopt^X))$.

\emph{Proof}: We defer the proof and a nontrivial example to Supplementary Material~\cite{SM}.

In our generalized thermostatistic formalism,
the solutions to \eref{eq:T2} give the most general form of the equilibrium distribution and
\eref{eq:dist} provides a recipe for finding it for the cases satisfying \eref{eq:maps}.
Solutions to \eref{eq:maps} can be guessed for a number of cases of
practical interest, as shown below. However, the analytical forms of $\FC$ and $\HC$ are expected to be difficult to find, in general.
Nevertheless, we demonstrate below a systematic numerical method that can find $\FC$ and $\HC$ for a given $\CB$ and $\HB$, and thus determine
the corresponding equilibrium thermostatistics in the general case.

\section{MaxEnt principle with correlations}
We now show that the MaxEnt principle for $S^{\text{GS}}$ gives an independent derivation of \eref{eq:T2} when the self-similar correlations are treated as prior data along with the normalization and mean energy conditions \protect\cite{Shore1980,Presse2013a}.
For composite system $AB$, the constraints for the normalization and mean energy are the conventional ones, i.e. $I(\{w^{AB}_{\z_{AB}}\})=\sum_{\z_{AB}} w^{AB}_{\z_{AB}} - 1=0$ and $E(\{w^{AB}_{\z_{AB}}\})=\sum_{\z_{AB}} w^{AB}_{\z_{AB}} H^{AB}_{\z_{AB}}- \bar{H}^{AB}=0$, respectively, where 
$\bar{H}^{AB}$ is the average energy.
The prior knowledge of the self-similar correlations is represented by \eref{eq:two-variable1} as a \emph{functional constraint} over the phase space.
Thus, the constrained maximization of $\SS^{\text{BGS}}(\{w^{AB}_{\z}\})/k$ leads to
\begin{align}  \label{eq:MaxEnt de}
  &\frac{\partial}{\partial w^{AB}_{\z_{AB}'}} \big[\!-\!\sum_{\z_{AB}} \ln w^{AB}_{\z_{AB}}
      \!+\!a I(\{w^{AB}_{\z_{AB}}\})\!+\!b E(\{w^{AB}_{\z_{AB}}\})\notag\\
      &\quad +\! \sum_{\z_{AB}} c_{\z_{AB}} ( w^{AB}_{\z_{AB}} - \CB(w^{A}_{\z_{A}},w^{B}_{\z_{B}}) ) \big]
      = 0
\end{align}
with Lagrange multipliers $a$, $b$, and $\{c_{\z_{AB}}\}$, where $c_{\z_{AB}}$ is a function over the phase space.

In~\cite{SM}, we show \eq{eq:MaxEnt de} yields
\begin{align}  \label{eq:ln C(w,w)-c = H(ln w - c, ln w -c)}
       &\frac{1}{b^{AB}}[\ln \CB(w^{A}_{\z_{A}},w^{B}_{\z_{B}})\!-\!a^{AB}\!-\!c^{AB}_{\z_{AB}}] \\
       &\quad =\HB(\frac{1}{b^{A}}[\ln w^{A}_{\z_{A}}\!-\!a^{A}\!-\!c^{A}_{\z_{A}}], \frac{1}{b^{B}}[\ln w^{B}_{\z_{B}}\!-\!a^{B}\!-\!c^{B}_{\z_{B}}]) ,\notag
\end{align}
where equilibrium distributions are given by $\ln w^X_{\z_X} = a^X+b^X H^X_{\z_X}+c^X_{\z_X}$ for phase space functions $c^{AB}_{\z_{AB}}$, $c^{A}_{\z_{A}}$, and $c^{B}_{\z_{B}}$ that satisfy the above equation.
Setting
$\GC^{-1}(w^X_{\z_{X}}) = \frac{1}{b^{X}}[\ln w^{X}_{\z_{X}}\!-\!a^{X}\!-\!c^{X}_{\z_{X}}]$
shows that \eq{eq:ln C(w,w)-c = H(ln w - c, ln w -c)} is equivalent to \eq{eq:T2}, and so the solutions found here are equivalent to those given by the solutions of \eqs{eq:w=w(H)} and (\ref{eq:T2}) for corresponding values of the Lagrange multipliers $a^X$ and $b^X$.

\section{Relationship with previously-studied thermostatistic classes}
\tref{table:main} compares the forms of $\FC$ and $\HC$, 
and limitations of various classes of distributions.

An interesting result is that, although the Tsallis distribution, $\frac{1}{Z_q}e_q^{-\beta_qH}$,
is known to exhibit nonadditive average energy~\protect\cite{Tsallis2009}, our formalism shows that
it corresponds to systems with \emph{additive} Hamiltonians
as demonstrated in the table.
Evidently, the nonadditivity of the average energy is due to correlations forming between subsystems
(see also~\protect\cite{SaadatmandPrep}).
Nevertheless, our results
effectively rule out the validity of
the Tsallis distribution for systems with an interaction term in the
Hamiltonian and satisfying
Criteria~I and II. 
This is also true for the examples of
multifractal and $\phi$-exponential-class thermostatistics~\protect\cite{Naudts2011,Tsallis2009}
that are characterized by the Tsallis distribution.

Moreover, our formalism
covers thermostatistics of extreme cases of
correlations, most notably the well-studied case of one-dimensional Ising ferromagnets
at vanishing temperature. See~\cite{SM} for details.
Aside from such trivial maximally-correlated cases and the long-range Ising models~\citep{Dauxois2002,Latora2002,Salazar1999,Salazar2000,Portesi1995,Apostolov2009} (corresponding to the \emph{third} row of Table I as long as subsystems are macroscopic), we are
not aware of any previous thermostatistic formalism that can describe nontrivial long-range-interacting
systems, as in the \emph{last} row, by finding the equilibrium distribution.

\section{Numerical example}
The well-studied examples discussed above all have analytic solutions. Next, we demonstrate the versatility of our approach by numerically evaluating the statistics of a complex long-range-interacting model (an extreme case of correlations and nonadditivity). 
To demonstrate how our approach might handle a practical problem, we intentionally choose composition rules,
\begin{align}
  \CB( w^A, w^B)=& w^A w^B\frac{(3.3- w^A)(3.3- w^B)}{2.3^2},
  \label{eqn:CEx}
  \\
  \HB(H^A,H^B)=& 0.7(H^A+H^B)\;,
  \label{eqn:HEx}
\end{align}
which have \emph{no} known analytical solution for $w$ within our formalism, and which would require extremely long-range interactions for the energy composition rule.

\begin{figure}[bth]
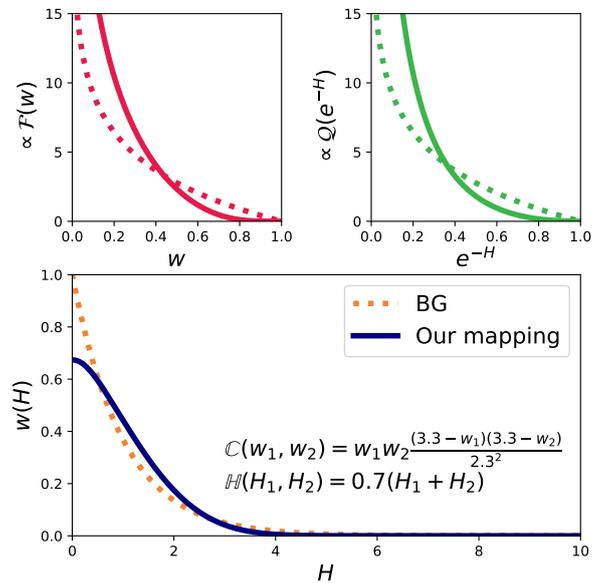

  \includegraphics[width=0.91\linewidth]{{{FigureForPaper}}}
  \caption{The top two panels show the mapping functions
    $\FC$ and $\QC$ for $\bar{w}$ and $q=e^{-H}$, respectively
    (solid lines), and their BG equivalents (dotted lines).
    The bottom panel shows
    the distribution $w(H)=\FC^{-1}(\beta(\HC(H) - \Hopt))$ \blk (solid line)
    for the mappings in Eqs.~(\ref{eqn:CEx}) and (\ref{eqn:HEx}),
    and the normalized BG distribution with the
    same average energy (dotted line).
    \label{fig:Mapping}}
\end{figure}

In \fref{fig:Mapping}, we show $\FC$ and $\QC$, where
$\HC(H)=\QC(e^{-H})$ [in BG, $\FC\propto\QC\propto\ln(x)$], for
Eqs.~\eqref{eqn:CEx} and \eqref{eqn:HEx}, and
$w$ versus $H$, where
$w(H)=\FC^{-1}(\beta(\HC(H) - \Hopt))$.
Here, $\Hopt$ ensures normalization
$\int  w(H)dH=\bar{1}=1$ and $\beta=-1$ ensures
$\int H \bar{w}(H)dH=\bar{H}=1$, which corresponds, in BG thermostatistics, to having
an inverse temperature of
$\beta^{\text{BG}}=-1$ \blk in unitless parameters.
It is interesting to see the
significant differences between the generalized distribution
$\bar{w}(H)$ and the normalized $w^{\text{BG}}=e^{-H}$ (bottom plot), being flatter for small energies and decaying more rapidly for larger energies. Full details of the numerical implementation are discussed in the Supplementary Material~\cite{SM}.

\section{Conclusions}
We employed an approach based on Liouville's theorem for equilibrium conditions obeying a
thermodynamic limit and self-similarity criterion, 
to provide an alternative
derivation of consistent generalized thermostatistics for systems with correlations and
nonadditive Hamiltonians (this is in comparison to the conventional MaxEnt
formulations~\protect\cite{Jaynes1957,Jaynes1957a,Naudts2011,Tsallis2009}). In our formalism, the equilibrium distributions of such systems are fully characterized by $\GC$ in \eref{eq:w=w(H)} or by
$\{\FC,\HC\}$ maps in \eref{eq:maps} for the special cases.
Upon appropriate choices of $\{\FC,\HC\}$,
our generalized thermostatistic class recovers well-established families,
i.e.~the standard Jaynes and Tsallis $q$-thermostatistics
as demonstrated in \tref{table:main}. Interestingly, our formalism implies that, for systems satisfying our criteria, the latter family of thermostatistics can \emph{only} capture the thermodynamics of systems with
additive Hamiltonians.

Our extension of the MaxEnt principle 
with $S^{\text{GS}}$ to include self-similar correlations as priors, gives an independent derivation of the same equilibrium distributions
derived using Liouville's theorem. This independent derivation confirms
the central role of the MaxEnt principle applied to $S^{\text{GS}}$ as a basis
for statistical inference in thermostatistics~\protect\cite{Shore1980}.
Moreover, it also clarifies the controversy surrounding the
heuristic application of the MaxEnt principle to generalized entropy functionals,
such as the Tsallis entropy.
Our derivation of the Tsallis distribution from Liouville's theorem and the MaxEnt principle applied to
$\SS^{\text{GS}}$, with a self-similar correlation prior, provides it with the mathematical support it previously lacked.
Moreover, the fact that the Tsallis distribution does not satisfy Shore and Johnson's
independent-system axiom~\protect\cite{Shore1980,Presse2013,Presse2015,Tsallis2015}
is no longer a problem, because it satisfies our criterion for self-similar correlated systems, which is more general than Shore and Johnson's system independence.

It would be interesting to examine the thermodynamics of
low-dimensional long-range Ising-type
models~\cite{Dauxois2002,Latora2002,Salazar1999,Salazar2000,Portesi1995,Apostolov2009,Thouless1969,Dyson1971,Ruelle1968}, which exhibit phase transitions
under certain conditions~\cite{Thouless1969,Dyson1971,Ruelle1968}.
In the context of our formalism, such phase transitions
are driven by the set of control parameters given above as $\{a^X,b^X\}$ (and which include the temperature through a global function~\citep{SaadatmandPrep}).

\begin{acknowledgments}
  We thank Baris Bagci for useful discussions. This research was funded by the Australian Research Council Linkage Grant No. LP180100096. J.A.V.~acknowledges financial support from Lockheed Martin Corporation.
  E.~G.~C.~was supported by the Australian Research Council Future Fellowship FT180100317. We acknowledge the traditional owners of the land on which this work was undertaken at Griffith University, the Yuggera people.
\end{acknowledgments}

\bibliographystyle{apsrev4-1}

%

\let\section=\OldSection
\clearpage
\widetext
\begin{center}
\textbf{\large Supplemental material for ``Thermodynamics from first principles: correlations and nonextensivity''} \\ \bigskip
\normalsize S.~N.~Saadatmand, Tim Gould, E.~G.~Cavalcanti, and J.~A.~Vaccaro\normalsize
\end{center}
\setcounter{equation}{0}
\setcounter{figure}{0}
\setcounter{table}{0}
\setcounter{page}{1}
\makeatletter
\renewcommand{\theequation}{S\arabic{equation}}
\renewcommand{\thefigure}{S\arabic{figure}}
\renewcommand{\bibnumfmt}[1]{[S#1]}
\renewcommand{\citenumfont}[1]{S#1}

In this supplemental material, we first discuss the 
restrictions on the normalization of Tsallis distribution and when it can be considered as a valid probability.
We then present 
a simple example of finding the distribution function, $\GC$, satisfying Eq.~(8) of the
main text. The proof 
of Theorem II of the main text and an associated nontrivial example 
is given in Sec.~III. Later, in Sec.~IV, we discuss
how to find Eq.~(13) from Eq.~(12). In Sec.~V, we demonstrate
how our formalism recovers $w^{\rm BG}$ for the trivial case of one-dimensional Ising ferromagnets at vanishing temperatures. The details of our numerical approach to find $\FC$ and $\HC$ maps in Eq.~(9) is presented in the last section.

\section{Restrictions on the normalization of Tsallis distribution}

Care needs to be taken with the normalization of the 
Tsallis distribution, $w_\z\propto e_q^{\frak{b} H_\z}$, that is introduced following Eq.~(6) in the main text~\protect\cite{Lutsko2011}.  For example, it cannot be normalized and interpreted as a valid 
probability distribution when $q>1$, $\frak{b}<0$ and the 
Hamiltonian involves unbounded kinetic energy terms. 
It can, however, be normalised for $\mathfrak{b}>0$ for $q>\!2$.
Also, it is important to note that 
$\mathfrak{b}$ is not generally the negative of the inverse temperature 
and has a nontrivial connection to the physical temperature; a consistent equilibrium $q$-thermostatistics is presented in~\protect\cite{SaadatmandPrep}. 

\section{Recovering $w^{\rm BG}$ using Equation~(\ref{eq:T2})}

In the main text, we argued that finding $\GC$ to satisfy Eq.~(8) of the
main text results in the equilibrium distribution in Eq.~(3). 
Here is a simple example: consider a conventional 
short-range-interacting system where the relations $w^{AB} = w^A w^B$ and $H^{AB} = H^A + H^B$ hold to 
a very a good approximation. In this case,
$\CB(w^{A},w^{B}) = w^A w^B$ and $\HB(H^A, H^B) = H^A + H^B$.
Equation~(8) becomes $ w^{A} w^{B} = \GC(
H^{A} + H^{B} )$, which is trivially satisfied by any $\GC$ satisfying
$f_\z=\GC( a\ln(f_\z) + b )$ (or equivalently $\GC(H_\z)=e^{-b/a}e^{H_\z/a}$); here
$f_\z$ is a bounded, otherwise arbitrary, phase-space function, $\{a,b\}$ are constants of integration, 
and we have $b^{AB}=b^A+b^B$, where $b^X$ is proportional to the size of the system $X$ as before.
This gives the Boltzmann-Gibbs (BG) exponential class of distributions
$w^{\rm BG}_\z=e^{-\beta H_\z}/Z^{\rm BG}$ with $Z^{\rm BG}=\sum_z e^{-\beta H_z}=e^{-\beta\bar{H}+k^{-1}S^{\rm GS}}$, 
as expected.
Here, we have $\beta \equiv -1/a$, $b \equiv -\ln(Z^{\rm BG})/\beta = \bar{H}-S^{\rm GS}/(k\beta)$, $S^{\text{GS}}(\{w_{\z}\}) = -k \sum_{\z} w_{\z} \ln (w_{\z})$ and, therefore, it
is clear that the constants $\beta$, $\bar{H}$, and $Z^{\rm BG}$ can be generally interpreted as the inverse temperature of the equilibrium, internal mean energy, and partition function respectively (as illustrated in Table I of the main text).

\section{Proof of Theorem II}

We prove Theorem II  by combining  $\FC_X(\cdot)$ with the function $\GC_X(\cdot)$ in Eq.~(3) to give the composite function $\FC_X(w^X_{\z})=\FC\GC_{X''}(H^X_{\z})$ where $\FC\GC_{X''}\equiv\FC_X\circ\GC_{X'}$.  This allows Eq.~(9a), after $\CB(w^A,w^B)$ is replaced by $w^{AB}$ according to Eq.~(7b), to be written as
\begin{align}  \label{eq:FG}
    \FC\GC_{AB}(H^{AB})=\FC\GC_{A}(H^{A})+\FC\GC_{B}(H^{B}).
\end{align}
In comparison Eq.~(9b), with $\HB(H^A,H^B)$ replaced by $H^{AB}$ according to Eq.~(7a), is
\begin{align}  \label{eq:H}
    \HC_{AB}(H^{AB})=\HC_{A}(H^{A})+\HC_{B}(H^{B}),
\end{align}
which suggests that $\FC\GC_{X}(\cdot)$ and $\HC_X(\cdot)$ are related functions.  Indeed, the equality $\FC\GC_{X}(H^X)=\HC_X(H^X)$ satisfies \eqs{eq:FG} and \eqr{eq:H} as does the linear relationship
\begin{align}  \label{eq:FG=aH+b}
    \FC\GC_{X}(H^{X})=a^X\HC_X(H^X)+b^X
\end{align}
for constants $a^X$ and $b^X$ provided we adopt the system composition rules  
\begin{align}  \label{eq:a and b}
    a^{AB}=a^A=a^B,\qquad b^{AB}=b^{A}+b^{B}.
\end{align}
Applying the inverse function $\FC^{-1}$ to both sides of \eq{eq:FG=aH+b} and making use of Eq.~(3) then yields the desired result, Eq.~(10) with \eq{eq:a and b} as condition Eq.~(11).~$\Box$  

While other relationships may hold, the linear relationship in \eq{eq:FG=aH+b} and its corresponding equilibrium distribution in Eq.~(10) are sufficient for our purposes here.

As an example of the application of Theorem II, consider a nontrivial correlated and interacting system satisfying
$w^{AB} = \CB(w^{A},w^{B}) = w^A \otimes_q w^B$ and $H^{AB} = \HB(H^A, H^B) = H^A \oplus_p H^B$, where
$\otimes_q$ and $\oplus_p$ are the generalized product and sum of the $q$-algebra
respectively~\protect\cite{Suyari2006,Tsallis2009}.
It is clear that choosing $\FC=\ln_q$ will result in
$\FC(\CB( w^{A}, w^{B})) =  \FC( w^{A}) + \FC( w^{B})$, while
setting $\HC(H_{\z})=\ln(e_p^{H_{\z}})$ gives
$\HC(\HB(H^A, H^B)) = \HC(H^A) + \HC(H^B)$.
Therefore, Eq.~(7) tells us that the equilibrium distribution
is simply $w_\z= \exp_q\big(a\ln(e_p^{H_{\z}})+b\big)$. 
This example appears in Table I in the main text.

\section{Finding Equation~(13) given Equation~(12)}

In the main text, we argued that the constrained maximization of $S^{\text{GS}}(\{w^{AB}_{\z}\})/k$ leads to
$\frac{\partial}{\partial w^{AB}_{\z_{AB}'}} [-\sum_{\z_{AB}} \ln w^{AB}_{\z_{AB}}
       + a I(\{w^{AB}_{\z_{AB}}\})\!+\!b E(\{w^{AB}_{\z_{AB}}\})
       + \sum_{\z_{AB}} c_{\z_{AB}} ( w^{AB}_{\z_{AB}} - \CB(w^{A}_{\z_{A}},w^{B}_{\z_{B}}) )]
      = 0$
with Lagrange multipliers $a$, $b$, and $\{c_{\z_{AB}}\}$, where $c_{\z_{AB}}$ is a function over the phase space. 

As $\CB(w^{A}_{\z_{A}},w^{B}_{\z_{B}})$ has no explicit dependence on $w^{AB}_{\z_{AB}}$, 
the above equation results in $-\!1\!-\!\ln w^{AB}_{\z_{AB}}\!+\!a\!+\!bH^{AB}_{\z_{AB}}\!+\!c_{\z_{AB}}=0$.
From Axiom II, this gives
\begin{align}   \label{eq:ln w^x}
       \ln w^X_{\z_X} = a^X+b^X H^X_{\z_X}+c^X_{\z_X}
\end{align}
for $X=A$, $B$, and $AB$, where we have redefined $a^X$ as $1+a^X$ for convenience.
Taking \eq{eq:ln w^x} with $X=AB$, and substituting for $w^{AB}_{\z_{AB}}$ and $H^{AB}_{\z_{AB}}$ using Eq.~(6) gives
$\ln \CB(w^{A}_{\z_{A}},w^{B}_{\z_{B}}) = a^{AB}+b^{AB} \HB(H^A_{\z_{A}}, H^B_{\z_{B}})+c^{AB}_{\z_{AB}} $.
Taking this and substituting for $H^A_{\z_{A}}$ and $H^B_{\z_{B}}$ using \eq{eq:ln w^x} with $X=A$ and $B$, respectively, then yields
$\frac{1}{b^{AB}}[\ln \CB(w^{A}_{\z_{A}},w^{B}_{\z_{B}})\!-\!a^{AB}\!-\!c^{AB}_{\z_{AB}}] 
        =\HB(\frac{1}{b^{A}}[\ln w^{A}_{\z_{A}}\!-\!a^{A}\!-\!c^{A}_{\z_{A}}], \frac{1}{b^{B}}[\ln w^{B}_{\z_{B}}\!-\!a^{B}\!-\!c^{B}_{\z_{B}}])$ as required.

\section{Recovering $w^{\rm BG}$ for one-dimensional Ising ferromagnets}

Consider
macroscopic steady-state ground states of the 
nearest-neighbor Ising model,
$H_{\rm Ising} = - J \sum_i S_i S_{i+1},~J>0,~S_i=\pm1~\forall i$ (e.g.~see~\protect\cite{Baxter2007} for
a review). It can be easily shown that this system is
describable, with a good approximation, by
$w^{AB}_{(\mathfrak{i},\mathfrak{j})} = \CB(w^{A}_{\mathfrak{i}},w^{B}_{\mathfrak{j}}) =
w^A_{\mathfrak{i}}w^B_{\mathfrak{j}}\exp[-\gamma(1-|m^{AB}_{(\mathfrak{i},\mathfrak{j})}|)]$ and
$H^{AB}_{(\mathfrak{i},\mathfrak{j})} = \HB(H^A_{\mathfrak{i}}, H^B_{\mathfrak{j}}) = H^{A}_{\mathfrak{i}}+H^{B}_{\mathfrak{j}}$ (neglecting the interaction term and boundary effects due to the
large size of the systems) --- here,
$\mathfrak{i}$ denotes a collective spin state, $\gamma\rightarrow\infty$
(playing the role of the diverging inverse temperature and,
therefore, the exponential acts effectively as a $\delta_{m^{A}_{\mathfrak{i}},m^{B}_{\mathfrak{j}}}$-function).
Also, we use $m^{X}=\sum_i S_i/N^X$, where $N^X$ is the number of sites, to denote the magnetization per site for systems $X=A$, $B$,
and their composition $AB$.
(Notice that the self-similarity rules
already imply that, for all single systems, there are two highly likely, equiprobable,
and degenerate states with $m_{X}=\pm1$, as expected from the spontaneous magnetization.)
Similar to our previous BG-type example in Sec.~I above, it is easy to check
$H_{\mathfrak{i}}=\GC( a\ln(H_{\mathfrak{i}}) + \gamma(1-|m_{\mathfrak{i}}|)+b )$
satisfies Eq.~(8) for some constant $a$ and additive parameter $b$ 
(the middle term in $\GC$ argument always vanishes for single systems);
therefore, the equilibrium distribution is of
the $w^{\rm BG}$-form as expected.

\section{Detailed description of the numerical procedure}

Here, we detail the numerical procedure used to calculate
$\FC$ and $\HC$ for arbitrary mappings $\CB$ and $\HB$.
These procedures were used to generate Fig.~(1) in the
main text.

Of relevance here are
three primary points: 1) that $\FC$ can be found accurately
in most cases; 2) that $\HC$ can be found similarly by
transforming to $q=e^{-H}$, to get
$\QB(q^A,q^B)=e^{-\HB(-\log(q^A),-\log(q^B))}$, $\HC(H)=\QC(e^{-H})$,
and $\HC^{-1}=-\log(\QC^{-1})$; 3) that for the BG case, we have
$\FC^{\BG}\propto -\log( w)$, $\QC^{\BG}\propto -\log( w)$ and
$\GC^{\BG}=e^{-H}$ in dimensionless units with $\beta^{\BG}=1$.

\subsection{Finding {$\FC$} and {$\FC^{-1}$} given {$\CB$}}

To calculate $\FC$, we use an iterative procedure over the mapping
$\CB$. Specifically, we exploit the fact that
the mapping $\CB(w^A,w^B)$ has attractors for
$w^{A/B}=0$, and a non-attractive fixed point $w^A=w^B=1$,
and that these are the only fixed points in $[0,1]^2$).

Thus, we can use the following procedure:
\begin{enumerate}
\item Choose an initial weight value $w_0=0.99<1$, and set
  $\FC(w_0)=f_0=0.01$.
\item Choose a secondary value $w_1=\CB(w_0,w_0)$,
  so that $\FC(w_1)=\FC(w_0)+\FC(w_0)=2f_0$.
\item Iterate $w_{n>1}=\CB(w_{n-1},w_n)$ until $w_n<1E-5$,
  and evaluate $\FC(w_{n>1})=\FC(w_{n-1})+\FC(w_{n-2})$
  using existing values.
\end{enumerate}
This gives a set of pairs of values $(w_i,\FC(w_i))$
over $i$, where we got $O(10)$ pairs in all our tests.

Our next step is to generate a continuous function $\FC(w)$.
We recognize that, for the BG mapping $\CB_{\BG}(w^A,w^B)=w^Aw^B$,
we get $\FC_{\BG}(w)\propto-\log(w)$. Running the BG case through
the distribution gives $w_i=w_0^i$, and $\FC(w_i)=(i+1)f_0$.
Clearly, this gives evenly distributed pairs
$(\log(w_i),\FC(w_i))$.

We assume that this behaviour is approximately preserved in
general mappings. We thus evaluate $\FC(w)$ for general
$w$ by interpolating (using a cubic spline)
the pairs we obtained by iteration on
the logarithm of the weights, i.e., we interpolate
$\FC(w_n)$ versus $\log(w_n)$. This method is exact for the BG case.
Without loss of generality, we finally normalize $\FC$ so that
$\int_0^1x\FC(x)dx=1$.

As a final step, we recognise that $\FC$ is monotone. This means
we can similarly find the inverse function $\FC^{-1}(z)$ by
interpolating $\log(x_n)$ versus
$\FC(x_n)$, which is again exact for the BG case.

\subsection{Finding {$\HC$} and {$\HC^{-1}$} given {$\HB$}}

We note that this iterative approach does not work for $\HB$,
which does not have any fixed point. But it
does work for $q=e^{-H}$, giving
\begin{align}
  \QB(q^A,q^B)=&e^{-\HB(-\log(q^A),-\log(q^B))},
\end{align}
with $\HC(H)=\QC(e^{-H})$ and $\HC^{-1}=-\log(\QC^{-1})$. We can
thus use the above approach to calculate mappings for
$\HB$, by going via $\QB$. Note that in the BG case, we find
$\QC_{\BG}(q)\propto -\log(q)$ and see that the method is
once again exact.

\bibliographystyle{apsrev4-1}

\cleardoublepage
\end{document}